\documentclass[12pt]{article}
\usepackage[utf8]{inputenc}
\usepackage{amsmath,amssymb,amsfonts}
\usepackage{graphicx}
\usepackage{hyperref}
\usepackage{natbib}
\usepackage{xcolor}
\usepackage{booktabs}
\usepackage{float}
\usepackage{tabularx}
\usepackage{colortbl}
\usepackage[margin=1in]{geometry}
\usepackage[font=small,labelfont=bf,textfont=it,width=0.8\textwidth]{caption}

\title{Prediction decomposition for causal analysis}
\author{Ofir Reich\thanks{
\href{Independent Researcher, mailto:ofir.re@gmail.com}{\texttt{ofir.re@gmail.com}}}}
\date{\today}

\begin{document}

\maketitle

\begin{abstract}
There is rising interest in using Machine Learning (ML) model predictions as outcomes in causal analysis. However, these methods have faced challenges in finding the true treatment effects. It is also challenging to make choices about which prediction models to choose, since we are interested not only in the accuracy of the prediction but in its ability to produce the correct causal effect in the analysis. In this paper I propose a decomposition of the prediction into between-unit prediction ($\eta_\mu$), within-unit-across-time prediction ($\eta_\epsilon$), and counterfactual-treatment-effect prediction ($\eta_T$). I show that the counterfactual-treatment-effect component is the one that determines whether the model recovers the true treatment effect, but only the first two components can be estimated from non-experimental data. I argue that within-unit-across-time prediction accuracy ($\eta_\epsilon$) is a structurally better proxy for the counterfactual-treatment-effect component ($\eta_T$) than overall prediction accuracy, and propose a metric to estimate it from panel data with at least two time periods. This metric serves as a diagnostic and model-selection tool for choosing ML models for causal analysis. Under the stronger assumption that $\eta_T \approx \eta_\epsilon$, it also enables constructing an approximately unbiased estimate of the treatment effect. I develop the theoretical framework and illustrate it with simulations of synthetic data.
\end{abstract}

\section{Introduction}
There is rising interest in using Machine Learning (ML) model predictions as outcomes in causal analysis. This could be in a Randomized Controlled Trial (RCT) or other kinds of analysis. The advantage of using ML predictions is that they are generally easier and cheaper to generate for large samples than individual data collection. One prominent example is research studying the effects of Unconditional Cash Transfers on various social outcomes (e.g. consumption or wealth), where the outcome is predicted using mobile phone Call Detail Records (\cite{blumenstock2025Haiti_CDR_fails_in_impact_eval, blumenstock2025Togo_surveys_vs_digital_for_impact_eval}). Another example is studying the effects of agricultural interventions on yields, where yields are predicted using Remote Sensing data (\cite{Cole2025shawn_jessica_pxd_paper, burke2021using, lobell2020eyes}). And more (\cite{Ratledge2022Uganda_satellite_electrification_and_decompression}).

The general structure of these analyses is as follows:
\begin{enumerate}
\item The actual outcome of interest is collected for a subsample of data.
\item A Machine Learning (ML) model is trained on this subsample of data for which both the actual label (the outcome of interest) and the broadly-available feature data (e.g. Call Detail Records, or Remote Sensing data) are known.
\item The ML model is used to predict the outcome of interest for each randomization unit, especially those outside the labeled subsample.
\item The causal analysis then proceeds as usual, using these ML-predicted outcomes as the outcome in the analysis. (This is not to be confused with using Machine Learning methods for the causal analysis itself when all data are known, which is an unrelated topic).
\end{enumerate}

A natural question is why one cannot simply estimate the treatment effect using the labeled subsample directly, since if the labeled data are randomly sampled and treatment is randomly assigned, the labeled subset yields an unbiased estimate. The motivation for using ML-predicted outcomes is not identification but rather scale, cost, and statistical power: the labeled subsample is typically too small to detect treatment effects with adequate precision, while ML predictions can be generated cheaply for the entire sample (\cite{blumenstock2025Togo_surveys_vs_digital_for_impact_eval, Ratledge2022Uganda_satellite_electrification_and_decompression}). It is also worth noting that if there is selection into the labeling process---for example, if labels are easier to collect in certain areas or for certain types of units---non-random labeling could itself introduce bias, providing further motivation for extending to the full sample via ML predictions.

\begin{table}[H]
\centering
\begin{tabularx}{\textwidth}{@{}>{\raggedright\arraybackslash}X >{\raggedright\arraybackslash}X >{\raggedright\arraybackslash}X >{\raggedright\arraybackslash}X@{}}
\toprule
\textbf{Intervention} & \textbf{Outcome of \newline interest} & \textbf{ML feature data} & \textbf{Unit of analysis} \\
\midrule
Unconditional Cash Transfers & Consumption, wealth & Mobile phone Call Detail Records & Person \\
\midrule 
Agricultural extension & Yield & Remote sensing pictures & Plot \\
\bottomrule
\end{tabularx}
\caption{Examples of interventions studied using ML-predicted outcomes}
\label{tab:interventions}
\end{table}

However, these methods have faced challenges in finding the true treatment effects (\cite{blumenstock2025Haiti_CDR_fails_in_impact_eval, blumenstock2025Togo_surveys_vs_digital_for_impact_eval}). A true causal effect exists when using the actual outcome data, but not when using the ML-predicted outcome data (\cite{blumenstock2025Togo_surveys_vs_digital_for_impact_eval}). Or the distribution of the ML-predicted outcomes is compressed relative to the label distribution, biasing the estimated effect downwards (\cite{Ratledge2022Uganda_satellite_electrification_and_decompression}). This is naturally a problem for this approach, especially if it cannot be known in advance without collecting the ground-truth outcome data for the entire sample, which defeats the purpose of using ML-predicted outcomes. A related problem is choosing the best ML algorithm, the right features etc. - since we are interested not only in the accuracy of the prediction but in its ability to produce the right causal effect in the analysis.

This problem has a close conceptual parallel in the surrogate outcome literature in biostatistics and economics. A surrogate endpoint is a proxy variable used in place of a primary outcome, and the central question is whether the treatment effect on the surrogate recovers the treatment effect on the true outcome (\cite{prentice1989surrogate}). \citet{frangakis2002principal} formalize this using principal stratification in a potential-outcomes framework, and \citet{athey2025surrogate} develop a ``surrogate index'' for combining short-term proxies to estimate long-term treatment effects. In our setting, the ML-predicted outcome functions as a surrogate for the actual outcome, and the core concern---that the treatment effect on the prediction may not match the treatment effect on the true outcome---is the same. The decomposition we propose provides a concrete diagnostic for when this surrogate is adequate.

In this paper I propose a metric to measure whether the trained ML model is better or worse at producing outcomes which would reveal the true causal effect. It requires panel data of the same form described above, for at least two time periods. This allows making informed choices about the right model to generate the ML-predicted outcomes, and a criterion for telling whether it is not good enough. I illustrate the metric with simulations of synthetic data, and justify it with a theoretical framework.

\section{Intuition - fitting to between-unit differences}
The problem with using ML models to predict outcomes which are later used in causal analysis is that ML models only "care" about prediction, whereas causal inference asks a question about the counterfactual. More concretely, a ML-model could predict very accurately the observed outcome for an individual, but not necessarily the difference between the outcome with and without Treatment. The ML model fits (among other things) to the variation \textit{between} units, whereas causal inference is interested in (counterfactual) variation \textit{within} unit. So for example a Machine Learning model using Call Detail Records could learn to predict a person's consumption level by whether they live in a wealthier area, and get very good prediction results. But a Cash Transfer would not change the area of the person's house, so the model would predict a zero effect of Cash Transfers on consumption, even if such an effect exists. Another example is that a yield prediction model could use only the farmer's last-season yield to predict their current season's yield, have very good predictions but obviously will not be able to detect any treatment effect.

We want to know that our model does not only fit to between-unit variation, but also to the treatment effect, and at the very least to natural (not counterfactual) within-unit variation.

There is moreover a structural reason to expect that $\eta_{\epsilon}$ is a better proxy for $\eta_T$ than $\eta_{\mu}$. Between-unit variation---captured by $\eta_{\mu}$---is largely driven by stable, immutable characteristics of the unit: geographic location, demographic attributes, historical baseline levels. These are precisely the features that do \textit{not} change in response to a treatment. A Cash Transfer does not move a household to a wealthier neighborhood; an agricultural extension does not change a farm's soil type or elevation. A model that fits well to between-unit variation by learning these static features will, by construction, be largely insensitive to treatment-induced changes. By contrast, within-unit variation across time---captured by $\eta_{\epsilon}$---is driven by features that do change over time: recent activity patterns, seasonal signals, transient shocks. A model that must fit to this dynamic variation is responsive to signals of the same nature as a treatment effect. We therefore argue that $\eta_{\epsilon}$ is not merely a convenient proxy for $\eta_T$, but a structurally motivated one.

\section{Basic case: 2 time periods}
We shall start with the simple case where outcome data is collected for only two time periods, for a subsample of units. Suppose the setting is a RCT, and the units are people.

Suppose the actual outcome for a person $i$ in time $t$ with/without Treatment is modeled as:
\begin{equation}
\textrm{actualOutcome}_{i,t} = \alpha + \mu_i + \gamma \textrm{Treat}_{i,t} + \epsilon_{i,t}
\end{equation}
where:
\begin{itemize}
\item $\alpha$ is a constant intercept
\item $\mu_i$ are person fixed characteristics, with mean 0.
\item $\textrm{Treat}_{i,t}$ is a random Treatment indicator
\item $\epsilon_{i,t}$ is an independent error term
\item $t$, the time period, takes values 1 and 2.
\end{itemize}

Note that $\alpha$, $\mu_i$ and $\textrm{Treat}_{i,t}$ are pairwise uncorrelated. Likewise, $\epsilon_{i,1}$ and $\epsilon_{i,2}$ are uncorrelated, due to $\mu_i$ soaking up that correlation.

\subsection{Pathological case: ML predictions fit only to between-person variation}
Now suppose we trained a ML model only on untreated persons, and that the prediction fits only (and perfectly) to the person fixed characteristics.
\begin{equation}
\textrm{predictedOutcome}_{i,t} = \alpha + \mu_i
\end{equation}

The prediction could be good when measured between people (say, on a test set without experimental variation), and even have good R-squared. But when using it to measure treatment effect, we would try to run the regression:
\begin{equation}
(\alpha + \mu_i =)\textrm{predictedOutcome}_{i,t} = \xi_0 + \xi_1 \textrm{Treat}_{i,t} + \nu_{i,t}
\end{equation}

And since $\textrm{Treat}_{i,t}$ is independent of $\mu_i$ and so uncorrelated, we would estimate $\xi_1 = 0$, i.e. a treatment effect of 0.

A way to know that this is the case beforehand is to check if the ML-prediction explains within-person variation across time periods, even for non-treated persons (so $\textrm{Treat}_{i,t} = 0$).

In this pathological case we would find
\begin{align}
\textrm{predictedOutcome}_{i,2} - \textrm{predictedOutcome}_{i,1} &= \mu_i - \mu_i = 0\\
\textrm{actualOutcome}_{i,2} - \textrm{actualOutcome}_{i,1} &= (\epsilon_{i,2} - \epsilon_{i,1}) \neq 0
\end{align}

So we would find that the predicted outcome explains precisely 0 percent of the within-person variation in outcome. This will be the basis of the metric I develop.

\subsection{More realistic case: ML predictions fit partially to between-person variation}
We shall now generalize to less pathological cases, where the ML predictions fit partially to between-person variation, partially to within-person between-period variation and partially to within-person between-counterfactuals variation.

As before
\begin{equation}
\textrm{actualOutcome}_{i,t} = \alpha + \mu_i + \gamma \textrm{Treat}_{i,t} + \epsilon_{i,t}
\end{equation}

Again suppose we trained a ML model only on untreated persons. We now decompose the ML-predicted outcome (on the test set, still of untreated persons) into these same components, as follows:
\begin{equation}
\textrm{predictedOutcome}_{i,t} = \alpha + \eta_{\mu}\mu_i + \eta_{T}\gamma\textrm{Treat}_{i,t} + \eta_{\epsilon}\epsilon_{i,t} + \nu_{i,t}
\end{equation}
where
\begin{itemize}
    \item $\eta_{\mu}$ -- Indicates how well the model fits to between-unit variation. A value of 1 means the model perfectly captures differences between individuals, while 0 means it ignores these differences entirely.
    
    \item $\eta_{\epsilon}$ -- Measures how well the model fits to within-unit temporal variation. When $\eta_{\epsilon} = 1$, the model perfectly captures how individuals naturally change over time (unrelated to treatment), while $\eta_{\epsilon} = 0$ means the model misses this variation completely.
    
    \item $\eta_{T}$ -- Represents how well the model captures counterfactual treatment effects. An $\eta_{T}$ value of 1 means the model perfectly predicts the causal impact of treatment, while 0 indicates the model is completely insensitive to treatment effects.

    \item $\nu_{i,t}$ is an independent error term, uncorrelated with the other elements.
    
    \item We assumed that the ML model gets the mean of the population, $\alpha$, right.
\end{itemize}

Note that even though the ML model was not trained on Treated persons (so on a population without Treatment variation), it could still capture some or all of the counterfactual Treatment variation - for example a perfectly accurate predictor would capture all of it.

Now suppose that we use our ML predicted outcomes as the dependent variable in treatment effect regression, in the standard fashion.

\begin{equation}
\textrm{predictedOutcome}_{i,t} = \xi_0 + \xi_1\textrm{Treat}_{i,t} + \delta_{i,t}
\end{equation}

Using the decomposition of predicted outcomes above, and the fact that Treatment was randomly assigned and is uncorrelated with other elements, we can see that the estimated treatment effect would be:
\begin{equation}
\hat{\xi}_1 = \eta_T \cdot \gamma
\end{equation}
and not the desired true treatment effect, $\gamma$.

Ideally, therefore, we want the model with $\eta_{T} = 1$, and we would find it using a population with experimental variation and check which model predicts it best. But usually this is not observable because the models are trained on non-Treated populations, and if we had a large enough sample to know which model best predicts the treatment effect, we would probably have a large enough sample to just estimate the treatment effect from that population directly. In lieu of estimating $\eta_{T}$, I conjecture that $\eta_{\epsilon}$ is a better proxy for $\eta_{T}$ than the overall prediction accuracy, and therefore estimate $\eta_{\epsilon}$. Before we derive our method, we shall see the problem with using prediction accuracy as the model selection criterion.

\subsubsection{The problem with prediction accuracy}

Normally ML models are chosen based on their prediction accuracy on the test set (a random holdout from the labeled training set). Considering our decomposition of prediction, we can now see why this is inadequate. If we estimated the overall prediction performance of the ML model on a non-Treated population (similar to its training set), using R-squared, we would have, using the equations above and the pairwise uncorrelatedness:
\begin{align}
\textrm{Cov}[\textrm{predictedOutcome}_{i,t}, \textrm{actualOutcome}_{i,t}] &= \eta_{\mu}\textrm{Var}[\mu_i] + \eta_{\epsilon}\textrm{Var}[\epsilon_{i,t}]
\end{align}

\begin{align}
R^2 &= \frac{\textrm{Cov}[\textrm{predictedOutcome}_{i,t}, \textrm{actualOutcome}_{i,t}]^2}{\textrm{Var}[\textrm{predictedOutcome}_{i,t}]\textrm{Var}[\textrm{actualOutcome}_{i,t}]}\\
&= \frac{(\eta_{\mu}\textrm{Var}[\mu_i] + \eta_{\epsilon}\textrm{Var}[\epsilon_{i,t}])^2}{(\eta_{\mu}^2\textrm{Var}[\mu_i] + \eta_{\epsilon}^2\textrm{Var}[\epsilon_{i,t}] + \textrm{Var}[\nu_{i,t}])(\textrm{Var}[\mu_i] + \textrm{Var}[\epsilon_{i,t}])}.
\end{align}

This depends on the values of $\eta$, but also on the relative size of the variances, $\textrm{Var}[\mu_i]$, $\textrm{Var}[\epsilon_{i,t}]$, $\textrm{Var}[\nu_{i,t}]$. If $\textrm{Var}[\mu_i] \gg \textrm{Var}[\epsilon_{i,t}]$, then $\eta_{\epsilon}$ will not matter much for the R-squared. We shall see this in simulations later.

\subsubsection{Estimating $\eta_{\epsilon}$}
\label{sec:estimating_eta_epsilon}
We shall now define and calculate a few quantities that will help us estimate $\eta_\epsilon$.
Define the difference across time periods within person for any variable $X_{i,t}$ as 
\begin{equation}
\Delta X_i = X_{i,t=2} - X_{i,t=1}.
\end{equation}

Calculating this difference for actualOutcome we get:
\begin{equation}
\Delta\textrm{actualOutcome}_i = \gamma\Delta\textrm{Treat}_i + \Delta\epsilon_i
\end{equation}

and similarly for $\Delta\textrm{predictedOutcome}_i$:
\begin{align}
\Delta\textrm{predictedOutcome}_i &= \textrm{predictedOutcome}_{i,2} - \textrm{predictedOutcome}_{i,1}\\
&= \eta_{T}\gamma\Delta\textrm{Treat}_i + \eta_{\epsilon}\Delta\epsilon_i + \Delta\nu_i
\end{align}

The left-hand-side quantities are observable for our subsample of persons with collected outcomes in both periods. We can see how well the actual outcome predicts the predicted outcome, in terms of linear regression.
\begin{align}
\textrm{Cov}[\Delta\textrm{predictedOutcome}_i, \Delta\textrm{actualOutcome}_i] &= \eta_{T}\gamma^2\textrm{Var}[\Delta\textrm{Treat}_i] + \eta_{\epsilon}\textrm{Var}[\Delta\epsilon_i]\\
&= \eta_{T}\gamma^2\textrm{Var}[\Delta\textrm{Treat}_i] + \eta_{\epsilon}\textrm{Var}[\Delta\epsilon_i]
\end{align}

But observe that
\begin{align}
\textrm{Var}[\Delta\epsilon_i] &= \textrm{Var}[\epsilon_{i,2} - \epsilon_{i,1}]\\
&= \textrm{Var}[\epsilon_{i,2}] + \textrm{Var}[\epsilon_{i,1}]\\
&= 2\textrm{Var}[\epsilon_{i,t}]
\end{align}
where the next-to-last equality is because errors are uncorrelated across periods, due to $\mu_i$ soaking up that correlation.

So finally we have
\begin{equation}
\textrm{Cov}[\Delta\textrm{predictedOutcome}_i, \Delta\textrm{actualOutcome}_i] = \eta_{T}\gamma^2\textrm{Var}[\Delta\textrm{Treat}_i] + \eta_{\epsilon}2\textrm{Var}[\epsilon_{i,t}]
\end{equation}

If we are working with a subsample of non-Treated persons, this simplifies even further.
\begin{align}
\textrm{Cov}[\Delta\textrm{predictedOutcome}_i, \Delta\textrm{actualOutcome}_i] &= \eta_{\epsilon}2\textrm{Var}[\epsilon_{i,t}]\\
\textrm{Var}[\Delta\textrm{actualOutcome}_i] &= 2\textrm{Var}[\epsilon_{i,t}]\\
\end{align}

If we ran a linear regression of $\Delta\textrm{predictedOutcome}_i$ on $\Delta\textrm{actualOutcome}_i$, without an intercept:
\begin{equation}
\Delta\textrm{predictedOutcome}_i = \beta \Delta\textrm{actualOutcome}_i + \textrm{error}_i,
\end{equation}

the estimated coefficient is 
\begin{equation}
\hat{\beta} = \frac{\textrm{Cov}[\Delta\textrm{predictedOutcome}_i, \Delta\textrm{actualOutcome}_i]}{\textrm{Var}[\Delta\textrm{actualOutcome}_i]} = \frac{\eta_{\epsilon}2\textrm{Var}[\epsilon_{i,t}]}{2\textrm{Var}[\epsilon_{i,t}]} = \eta_{\epsilon}
\end{equation}

This is the parameter we are interested in estimating, in absence of a direct estimate of $\eta_{T}$.

This empirical diff-vs.-diff slope will be our estimate of $\eta_{\epsilon}$, and the standard error from the regression will be our standard error for the estimate.

\subsubsection{Estimating $\eta_{\mu}$}
In a non-Treated population, doing some algebra using the definitions and the uncorrelatedness of the different components, we have:
\begin{align}
\textrm{Cov}[\textrm{predictedOutcome}_{i,t}, \textrm{actualOutcome}_{i,t}] &= \eta_{\mu}\textrm{Var}[\mu_i] + \eta_{\epsilon}\textrm{Var}[\epsilon_{i,t}]\\
\textrm{Cov}[\Delta\textrm{predictedOutcome}_i, \Delta\textrm{actualOutcome}_i] &= \eta_{\epsilon}2\textrm{Var}[\epsilon_{i,t}]\\
\textrm{Var}[\textrm{actualOutcome}_{i,t}] &= \textrm{Var}[\mu_i] + \textrm{Var}[\epsilon_{i,t}]\\
\textrm{Var}[\Delta\textrm{actualOutcome}_i] &= 2\textrm{Var}[\epsilon_{i,t}]
\end{align}

We can perform some algebra to isolate $\eta_{\mu}$, where for ease of notation we replace predictedOutcome with predOut and actualOutcome with actualOut:
\begin{align}
\textrm{Var}[\textrm{actualOut}_{i,t}] - \frac{1}{2}\textrm{Var}[\Delta\textrm{actualOut}_i] &= \textrm{Var}[\mu_i]
\end{align}
\begin{align}
\textrm{Cov}[\textrm{predOut}_{i,t}, \textrm{actualOut}_{i,t}] - \frac{1}{2}\textrm{Cov}[\Delta\textrm{predOut}_i, \Delta\textrm{actualOut}_i] &= \eta_{\mu}\textrm{Var}[\mu_i]\\
\Rightarrow \frac{\textrm{Cov}[\textrm{predOut}_{i,t}, \textrm{actualOut}_{i,t}] - \frac{1}{2}\textrm{Cov}[\Delta\textrm{predOut}_i, \Delta\textrm{actualOut}_i]}{\textrm{Var}[\textrm{actualOut}_{i,t}] - \frac{1}{2}\textrm{Var}[\Delta\textrm{actualOut}_i]} &= \frac{\eta_{\mu}\textrm{Var}[\mu_i]}{\textrm{Var}[\mu_i]} = \eta_{\mu}
\end{align}

The left-hand side of the last equation will be our estimate for $\eta_{\mu}$. Standard errors can be estimated using bootstrap.

\subsubsection{Estimating $\eta_{T}$}
This is done by estimating the Treatment Effect twice, with a standard treatment effect regression. Once using the actual outcome, and once using the predicted outcome. We obviously need experimental variation for this, which we did not need to estimate the previous two coefficients, which makes this approach infeasible in most practical applications.

The specification
\begin{equation}
\textrm{actualOutcome}_{i,t} = \alpha + \mu_i + \gamma\textrm{Treat}_{i,t} + \epsilon_{i,t}
\end{equation}

(If is $\textrm{Treat}_{i,t}$ is fixed within unit, $\textrm{Treat}_{i,t} = \textrm{Treat}_i$, then the regression is run without the Fixed-Effects $\mu_i$)

Gives us the estimate $\hat{\gamma}_{\textrm{actual}} \rightarrow \gamma$.

Running the same regression with predictedOutcome as the left-hand side gives us the estimate $\hat{\gamma}_{\textrm{predicted}} \rightarrow \eta_{T} \gamma$. So our estimate for $\eta_{T}$ will be:
\begin{equation}
\hat{\eta}_{T} = \frac{\hat{\gamma}_{\textrm{predicted}}}{\hat{\gamma}_{\textrm{actual}}}
\end{equation}

Here, too, standard errors can be estimated using bootstrap.

\section{Simulations}
\subsection{Simulation process}
I use the framework above to simulate synthetic data and run several different tests. All code used to generate these simulations and charts in the paper is publicly available on GitHub (\cite{reich2025prediction_decomposition}).

\subsubsection{Simulating actual outcomes}
Recall our notation:
\begin{equation}
\textrm{actualOutcome}_{i,t} = \alpha + \mu_i + \gamma\textrm{Treat}_{i,t} + \epsilon_{i,t}
\end{equation}

I simulate actual outcomes for each person for two time periods. I selected the standard deviation of $\mu$ relative to the standard deviation of $\epsilon$ according to the results in a real setting which is not part of this paper (\cite{Cole2025shawn_jessica_pxd_paper}). In that setting the share of variance explained by person fixed effects was 0.92, so I used this value. Notice that this is a large share of the variance, but it comes from real data and is not necessarily atypical.

Specifically, I used values:
\begin{align}
\alpha &= 3200 \\
\textrm{SD}(\mu) &= 1400\\
\textrm{SD}(\epsilon) &= 600\\
\gamma &= 200\\
P(\textrm{Treat}_{i,t}) &= P(\textrm{Treat}_{i}) = 0.5
\end{align}

Treatment was fixed within person across time, so $\textrm{Treat}_{i,t} = \textrm{Treat}_i$.

We can then draw random numbers to obtain simulated values of $\mu_i$, $\epsilon_{i,t}$, $\textrm{Treat}_{i,t}$ for each person in each time period, and add them up to calculate the actual outcome for each person and time period using the formula for $\textrm{actualOutcome}_{i,t}$ above. I used a Log-Normal distribution for $(\alpha + \mu_i)$ to simulate the fat-tailed distribution of between-person outcomes present in many real-world settings.

I simulate those outcomes once, and hold them fixed when simulating the predicted outcomes explained below.

\subsubsection{Simulating predicted outcomes}
Recall our decomposition for predicted outcomes:
\begin{equation}
\textrm{predictedOutcome}_{i,t} = \alpha + \eta_{\mu}\mu_i + \eta_{T}\gamma\textrm{Treat}_{i,t} + \eta_{\epsilon}\epsilon_{i,t} + \nu_{i,t}
\end{equation}

I simulate the predicted outcomes as follows. I iterate over parameter values (between 0 and 1 in skips of 0.25) for each of $\eta_{\mu}$, $\eta_{T}$, $\eta_{\epsilon}$, and values (between 0 and 1,000 in skips of 250) for $\textrm{Var}[\nu]$. I used the full cartesian product of the parameter values, so each combination of values is obtained. So each simulation has a set of parameter values for $\eta_{\mu}, \eta_{T}, \eta_{\epsilon}, \textrm{Var}[\nu]$. For each simulation, I generate $\nu_{i,t}$ for each person and period, and calculate the predicted outcomes for each person and period using their values of $\mu_i$, $\epsilon_{i,t}$, $\textrm{Treat}_{i,t}$, $\nu_{i,t}$, and the values of $\eta_{\mu}$, $\eta_{T}$, $\eta_{\epsilon}$ by substituting in the formula for $\textrm{predictedOutcome}_{i,t}$ above.

\subsubsection{Calculating stats}
We calculate various stats:
\begin{itemize}
\item We run a linear regression of $\textrm{actualOutcome}_{i,t}$ on $\textrm{predictedOutcome}_{i,t}$ (without an intercept). We estimate:
  \begin{itemize}
  \item "ML prediction R-squared": R-squared of this regression.
  \end{itemize}
\item We run a linear regression of $\Delta\textrm{actualOutcome}_i$ on $\Delta\textrm{predictedOutcome}_i$ (without an intercept).
  \begin{equation}
  \Delta\textrm{predictedOutcome}_i = \beta\Delta\textrm{actualOutcome}_i + \textrm{error}_i 
  \end{equation}
  We estimate:
  \begin{itemize}
  \item The slope, i.e. the coefficient estimated for $\Delta\textrm{predictedOutcome}_i$.
  \item "Diff prediction R-squared": R-squared of this regression.
  \end{itemize}
\item Compression: ratio $\textrm{StD}[\textrm{predictedOutcome}]/\textrm{StD}[\textrm{actualOutcome}]$.
\item Share of variance in \textit{actualOutcome} explained by person fixed-effects $\mu_i$ (calibrated to 0.92). Note that this is a parameter we have chosen for the simulations, but it comes from a real setting, and it holds in many real settings - the treatment effect is almost always small compared with between-unit variation. This is why RCTs often spend time and effort collecting pre-intervention outcomes and perform a difference-in-differences analysis, since it cancels precisely that variation between units.
\end{itemize}

We then run an ordinary treatment effects linear regression using those \textit{predictedOutcomes} as the target variable:
\begin{equation}
\textrm{predictedOutcome}_{i,t} = \xi_0 + \xi_1\textrm{Treat}_{i,t} + \delta_{i,t}
\end{equation}

We estimate the Treatment Effect, $\hat{\xi}_1$.

We also estimate the actual Treatment Effect (which is different than the expectation due to statistical noise in the simulation) using the actual outcome, which is unknown in real settings where actualOutcome was not collected for the entire sample:
\begin{equation}
\textrm{actualOutcome}_{i,t} = \gamma_0 + \gamma_1\textrm{Treat}_{i,t} + \delta_{i,t}
\end{equation}

We calculate two other stats from this regression:
\begin{itemize}
\item "Scaled Treatment Effect": estimated slope divided by the actual Treatment Effect: $\hat{\xi_1}/\hat{\gamma_1}$.
\item t-statistic for the slope coefficient.
\end{itemize}

\subsection{Simulation results}
I find the following.

\subsubsection{Better ML-prediction does not guarantee more accurate treatment estimation.}
\begin{figure}[H]
\centering
\includegraphics[width=0.8\textwidth]{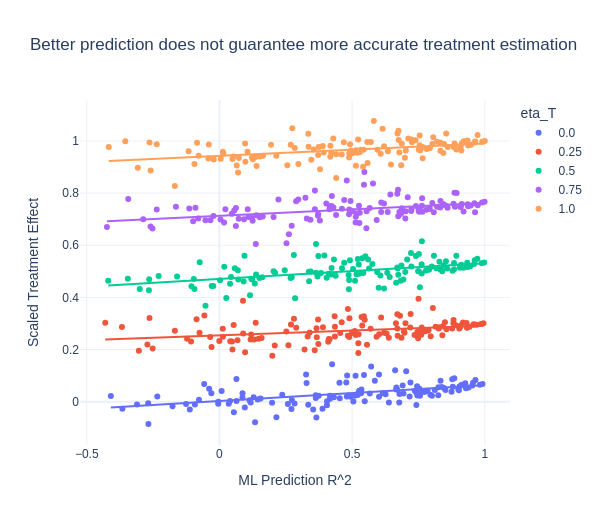}
\caption{Each dot is a specific simulation of actual outcomes and predicted outcomes. The x-axis marks the R-squared of the ML prediction, and the y-axis the Scaled Treatment Effect (where 1 is the correct effect). The color of the dot is by $\eta_{T}$. The trendline is for all points, using OLS. We can see the main determinant of the Scaled Treatment Effect is $\eta_{T}$, where the general prediction R-squared matters very little.}
\label{fig:better_prediction_not_better_treat_estimation}
\end{figure}

There could be two different models, one of which would have vastly inferior R-squared in predicting actual outcome, but would be better at giving the true treatment effect. See Figure~\ref{fig:better_prediction_not_better_treat_estimation}.

\subsubsection{Better ML-prediction is mostly determined by person attributes}
\begin{figure}[H]
\centering
\includegraphics[width=0.6\textwidth]{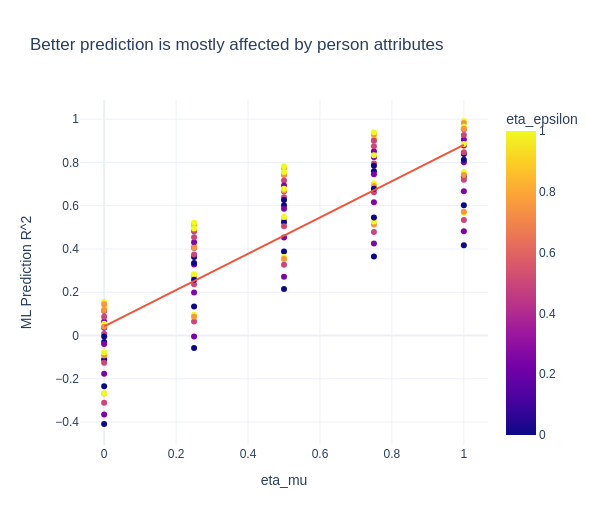}
\caption{Each dot is a simulation. X-axis is $\eta_{\mu}$. Y-axis is ML prediction R-squared. Trendline is OLS. Higher $\eta_{\mu}$ strongly predicts higher prediction R-squared.}
\label{fig:better_prediction_mostly_person_attributes}
\end{figure}

\begin{figure}[H]
\centering
\includegraphics[width=0.6\textwidth]{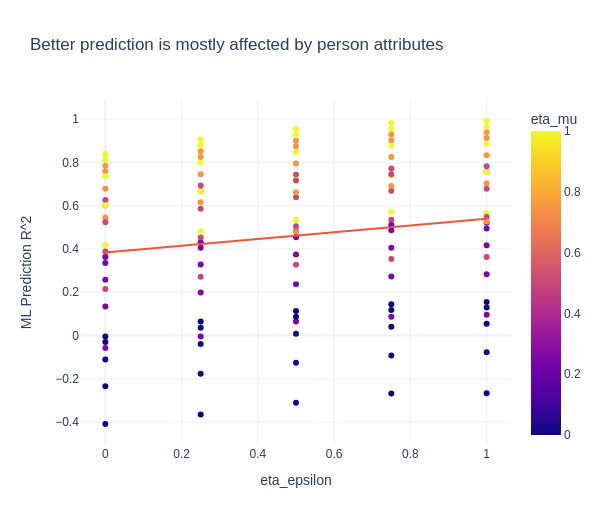}
\caption{Each dot is a simulation. X-axis is $\eta_{\epsilon}$. Y-axis is ML prediction R-squared. Trendline is OLS. $\eta_{\epsilon}$ and prediction R-squared don't have a strong relationship.}
\label{fig:better_prediction_mostly_person_attributes_2}
\end{figure}

In our setting, where $\textrm{StD}[\mu] \gg \textrm{StD}[\gamma\textrm{Treat}]$, better prediction is mostly about capturing the person fixed effects, and so $\eta_{T}$ does not greatly affect the prediction accuracy. See Figures~\ref{fig:better_prediction_mostly_person_attributes} and \ref{fig:better_prediction_mostly_person_attributes_2}.

\subsubsection{Better ML-prediction does not guarantee more statistical power for detecting treatment effect}
\begin{figure}[H]
\centering
\includegraphics[width=0.8\textwidth]{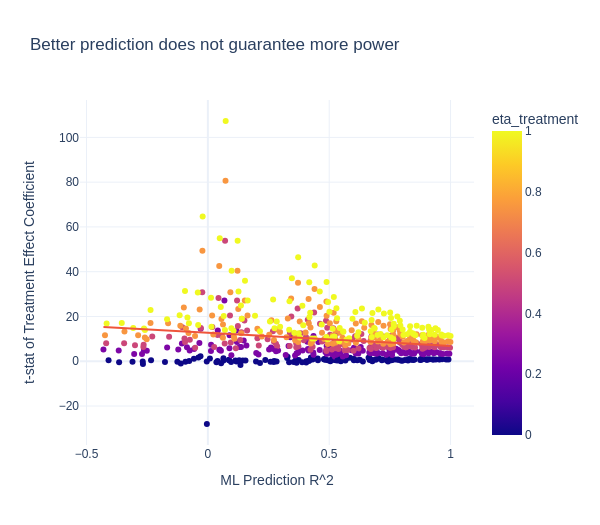}
\caption{Each dot is a simulation. X-axis is ML prediction R-squared. Y-axis is the t-statistic for the coefficient of Treat in the Treatment Effect regression. Trendline is OLS. The t-statistic is mostly determined by $\eta_{T}$, not by the prediction R-squared. A few very extreme outliers in t-statistic were discarded.}
\label{fig:better_prediction_not_more_power}
\end{figure}

See Figure ~\ref{fig:better_prediction_not_more_power}.

\subsubsection{Distribution compression of ML-Predicted outcomes is not predictive of treatment effect compression}
\begin{figure}[H]
\centering
\includegraphics[width=0.8\textwidth]{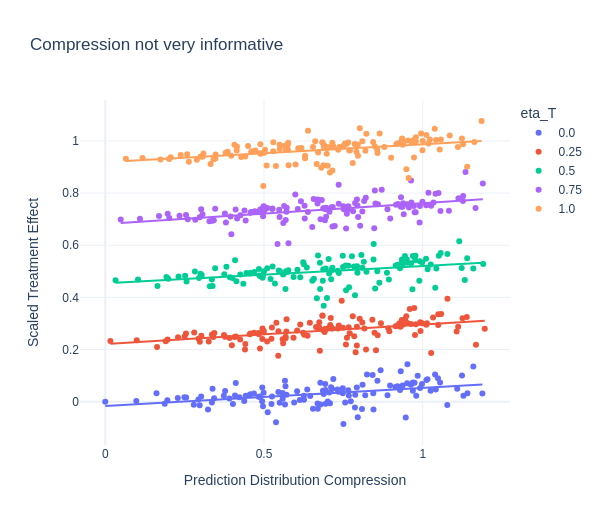}
\caption{Each dot is a simulation. X-axis is the compression ratio StD(predictedOutcome)/StD(actualOutcome). Y-axis is Scaled Treatment Effect. Trendline is OLS.}
\label{fig:compression_not_informative}
\end{figure}

This is for the same reason - the compression is mostly about $\eta_{\mu}$ where the correct estimated treatment effect is mostly affected by $\eta_{T}$. This is important since some papers (e.g. \cite{Ratledge2022Uganda_satellite_electrification_and_decompression}) have proposed to target the compression directly (and even artificially inflate the predictions) as a method for dealing with the treatment effect being compressed. This would only work if $\eta_{\mu}$ and $\eta_{T}$ are similar, since artificial inflation of the prediction inflates both. See Figure ~\ref{fig:compression_not_informative}.

More recently, \citet{pettersson2025onemap} propose to correct for prediction attenuation using Tweedie's formula or linear calibration, without requiring panel data or fresh ground-truth labels. While practically attractive, our framework helps explain why such corrections may be insufficient: they target the overall compression of predicted outcomes, which is primarily determined by $\eta_{\mu}$ (between-unit fit), whereas the scaled treatment effect is determined by $\eta_T$. Unless $\eta_{\mu} \approx \eta_T$---which need not hold, and which our structural arguments in Section 2 suggest is a poor assumption---correcting for overall compression will not in general recover the true treatment effect. The same critique applies to other compression-targeting approaches (\cite{prest2023rcts}). Our diff-vs-diff metric, by contrast, directly targets $\eta_{\epsilon}$, which we argue is structurally closer to $\eta_T$.

\subsubsection{Diff-vs-diff regression predicts the scaled treatment effect, when $\eta_{T} = \eta_{\epsilon}$}

\begin{figure}[H]
\centering
\includegraphics[width=0.8\textwidth]{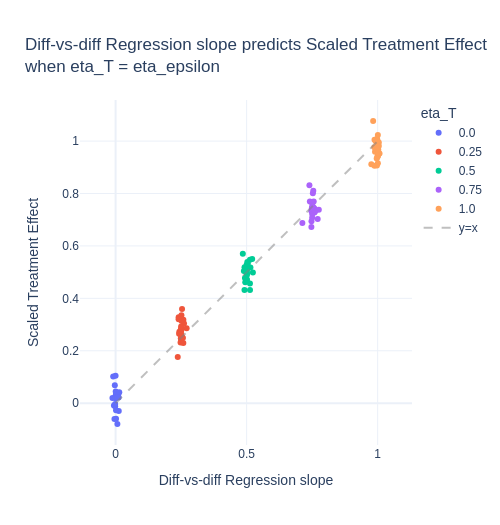}
\caption{Each dot is a simulation. X-axis is the diff-vs-diff regression slope (our estimate for $\eta_{\epsilon}$). Y-axis is Scaled Treatment Effect. Color is by $\eta_{T}$. Dashed line is y=x. When we restrict ourselves to cases where $\eta_{T} = \eta_{\epsilon}$, meaning the prediction fits to within-person variation as well as it fits to counterfactual treatment variation, then our method of estimating $\eta_{\epsilon}$ using the diff-vs-diff regression predicts the Scaled Treatment Effect rather well.}
\label{fig:diff_vs_diff_predicts_treat_effect}
\end{figure}

When we restrict ourselves to cases where $\eta_{T} = \eta_{\epsilon}$, meaning the prediction fits to within-person variation as well as it fits to counterfactual treatment variation, then our method of estimating $\eta_{\epsilon}$ using the diff-vs-diff regression predicts the Scaled Treatment Effect rather well. See Figure \ref{fig:diff_vs_diff_predicts_treat_effect}. This is to be expected, since the Scaled Treatment Effect is largely determined by $\eta_{T}$.

When our assumption that $\eta_{T} = \eta_{\epsilon}$ holds, we can correct the bias in the estimated Scaled Treatment Effect and arrive at an unbiased estimate of the Treatment Effect:
\begin{equation}
\textrm{UnbiasedTreatmentEffect} = \textrm{EstimatedTreatmentEffect} / \hat{\eta}_{\epsilon}
\end{equation}

\begin{figure}[H]
\centering
\includegraphics[width=0.8\textwidth]{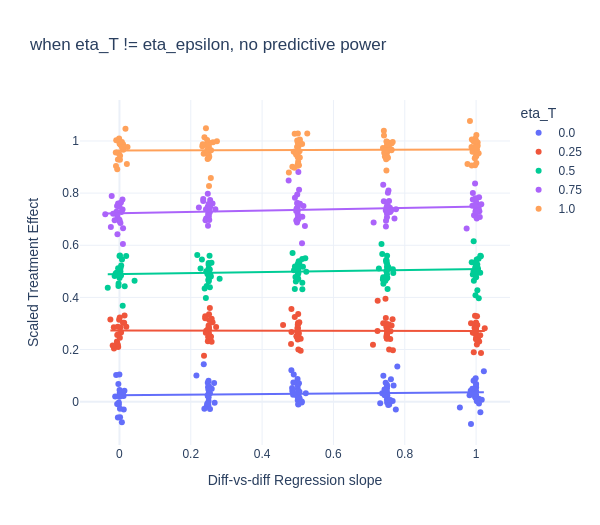}
\caption{Each dot is a simulation. X-axis is the diff-vs-diff regression slope (our estimate for $\eta_{\epsilon}$). Y-axis is Scaled Treatment Effect. Trend lines are OLS. With no restriction that $\eta_{T} = \eta_{\epsilon}$, there is no relationship between our estimate and the Scaled Treatment Effect.}
\label{fig:diff_vs_diff_all_points}
\end{figure}

Importantly, this is only true if we assume $\eta_{T} = \eta_{\epsilon}$. If there is no correlation between $\eta_{T}$ and $\eta_{\epsilon}$, obviously our regression slope aimed at estimating $\eta_{\epsilon}$ is not helpful at predicting the Scaled Treatment Effect affected by $\eta_{T}$. See Figure \ref{fig:diff_vs_diff_all_points}.

It is important to distinguish between two uses of $\hat{\eta}_{\epsilon}$:
\begin{itemize}
\item As a \textbf{diagnostic and model-selection tool}: comparing $\hat{\eta}_{\epsilon}$ across candidate ML models, or checking whether it is unacceptably low. This use is broadly valuable and does not require $\eta_{T} = \eta_{\epsilon}$ to hold exactly---it only requires that models with higher $\eta_{\epsilon}$ tend to have higher $\eta_T$, which is supported by the structural argument in Section 2.
\item As a \textbf{bias-correction factor}: dividing the estimated treatment effect by $\hat{\eta}_{\epsilon}$ to recover an unbiased estimate. This requires the stronger assumption that $\eta_T \approx \eta_{\epsilon}$, and should be applied with caution.
\end{itemize}

It remains an empirical question what is the typical relationship between $\eta_{\mu}$, $\eta_{T}$, $\eta_{\epsilon}$. I conjecture that in many contexts $\eta_{T}$ and $\eta_{\epsilon}$ will be related, at the very least more related than $\eta_{T}$ and $\eta_{\mu}$. But this remains to be studied in real settings, and it is possible that in practice some combination of $\eta_{\mu}$, $\eta_{\epsilon}$ will be a better predictor of $\eta_{T}$ than only $\eta_{\epsilon}$.

\section{Generalization to more than 2 time periods}

The generalization to more than 2 time periods is relatively straightforward. The decomposition remains the same, except now $t$ can take on more than 2 values, $t=1..T$.

\begin{align}
\textrm{actualOutcome}_{i,t} &= \alpha + \mu_i + \gamma\textrm{Treat}_{i,t} + \epsilon_{i,t}\\
\textrm{predictedOutcome}_{i,t} &= \alpha + \eta_{\mu}\mu_i + \eta_{T}\gamma\textrm{Treat}_{i,t} + \eta_{\epsilon}\epsilon_{i,t} + \nu_{i,t}
\end{align}

The generalization of the time-differences $\Delta X_i := X_{i,2}-X_{i,1}$ are centered variables, denoted by tildes:
\begin{equation}
\tilde{X}_{i,t} := X_{i,t} - \bar{X}_{i} = X_{i,t} - \frac{1}{T}\sum_{s=1}^{T}{X}_{i,s}
\end{equation}

\subsection{Estimating $\eta_{\epsilon}$}
Our regression specification using differences was:
\begin{equation}
\Delta\textrm{predictedOutcome}_i = \beta\Delta\textrm{actualOutcome}_i + \textrm{error}_i
\end{equation}

This is equivalent to the specification:
\begin{equation}
\textrm{predictedOutcome}_{i,t} = \alpha_i + \beta\textrm{actualOutcome}_{i,t} + \textrm{error}_{i,t}
\end{equation}

This fixed-effects regression extends unchanged to multiple time periods, so this will be our specification, and our estimate for $\eta_{\epsilon}$ remains $\hat{\beta}$. Standard errors for this estimate come directly from the regression.

\subsection{Estimating $\eta_{\mu}$}
This is very similar to the case of two time periods, except now we use not the deltas but the centered variables, $\tilde{X}_{i,t}$.

Here too we can perform some algebra to isolate $\eta_{\mu}$, where for ease of notation we replace predictedOutcome with predOut and actualOutcome with actualOut:
\begin{align}
\tilde{\textrm{actualOut}}_{i,t} &= \textrm{actualOut}_{i,t} - \overline{\textrm{actualOut}}_{i} = \gamma(\textrm{Treat}_{i,t} - \overline{\textrm{Treat}}_{i}) + \epsilon_{i,t} - \bar{\epsilon}_{i} = \gamma\tilde{\textrm{Treat}}_{i,t} + \tilde{\epsilon}_{i,t}\\
\tilde{\textrm{predOut}}_{i,t} &= \textrm{predOut}_{i,t} - \overline{\textrm{predOut}}_{i} = \eta_{T}\gamma\tilde{\textrm{Treat}}_{i,t} + \eta_{\epsilon}\tilde{\epsilon}_{i,t} + \tilde{\nu}_{i,t}
\end{align}

We'll now work out the Variance of $\tilde{\epsilon}_{i,t}$. First, we'll rewrite $\tilde{\epsilon}_{i,t}$:
\begin{align}
\tilde{\epsilon}_{i,t} &= \epsilon_{i,t} - \frac{1}{T}\sum_{s=1}^{T}\epsilon_{i,s} = \frac{T-1}{T}\epsilon_{i,t} - \frac{1}{T} \sum_{s\neq t}\epsilon_{i,s}
\end{align}

But since the errors in different time periods are assumed to be uncorrelated and have equal variance (if these assumptions do not hold, it requires a separate estimate which exceeds the bounds of this paper, but can perhaps be done using time-series methods):
\begin{align}
\textrm{Var}[\tilde{\epsilon}_{i,t}] &= \textrm{Var}[\epsilon_{i,t}]\left(\left(\frac{T-1}{T}\right)^2 + \left(\frac{1}{T}\right)^2 (T-1)\right) = \textrm{Var}[\epsilon_{i,t}]\frac{T-1}{T}
\end{align}

In a non-treated population:
\begin{align}
\textrm{Cov}[\tilde{\textrm{predOut}}_{i,t}, \tilde{\textrm{actualOut}}_{i,t}] &= \eta_{\epsilon}\textrm{Var}[\tilde{\epsilon}_{i,t}] = \eta_{\epsilon}\textrm{Var}[\epsilon_{i,t}]\frac{T-1}{T}\\
\textrm{Cov}[\textrm{predOut}_{i,t}, \textrm{actualOut}_{i,t}] &= \eta_{\mu}\textrm{Var}[\mu_i] + \eta_{\epsilon}\textrm{Var}[\epsilon_{i,t}]\\
\textrm{Var}[\textrm{actualOut}_{i,t}] &= \textrm{Var}[\mu_i] + \textrm{Var}[\epsilon_{i,t}]\\
\textrm{Var}[\tilde{\textrm{actualOut}}_{i,t}] &= \textrm{Var}[\tilde{\epsilon}_{i,t}] = \textrm{Var}[\epsilon_{i,t}]\frac{T-1}{T}
\end{align}

\begin{align}
\textrm{Cov}[\textrm{predOut}_{i,t}, \textrm{actualOut}_{i,t}] - \frac{T}{T-1}\textrm{Cov}[\tilde{\textrm{predOut}}_{i,t}, \tilde{\textrm{actualOut}}_{i,t}] &= \eta_{\mu}\textrm{Var}[\mu_i]\\
\textrm{Var}[\textrm{actualOut}_{i,t}] - \frac{T}{T-1}\textrm{Var}[\tilde{\textrm{actualOut}}_{i,t}] &= \textrm{Var}[\mu_i]
\end{align}

\begin{align}
\Rightarrow \frac{\textrm{Cov}[\textrm{predOut}_{i,t}, \textrm{actualOut}_{i,t}] - \frac{T}{T-1}\textrm{Cov}[\tilde{\textrm{predOut}}_{i,t}, \tilde{\textrm{actualOut}}_{i,t}]}{\textrm{Var}[\textrm{actualOut}_{i,t}] - \frac{T}{T-1}\textrm{Var}[\tilde{\textrm{actualOut}}_{i,t}]} = \frac{\eta_{\mu}\textrm{Var}[\mu_i]}{\textrm{Var}[\mu_i]} = \eta_{\mu}
\end{align}

And so the left hand side of this equation is our estimate for $\eta_\mu$. Standard errors can be estimated using bootstrap, similar to the two-period case.

\subsection{Estimating $\eta_{T}$}
As in the two-period case, estimating $\eta_{T}$ requires experimental variation. We run two regressions:

\begin{equation}
\textrm{actualOutcome}_{i,t} = \alpha + \mu_i + \gamma\textrm{Treat}_{i,t} + \epsilon_{i,t}
\end{equation}

This gives us the estimate $\hat{\gamma}_{\textrm{actual}} \rightarrow \gamma$. Running the same regression with predictedOutcome as the dependent variable gives us $\hat{\gamma}_{\textrm{predicted}} \rightarrow \eta_T \gamma$. Our estimate for $\eta_{T}$ is then:

\begin{equation}
\hat{\eta}_{T} = \frac{\hat{\gamma}_{\textrm{predicted}}}{\hat{\gamma}_{\textrm{actual}}}
\end{equation}

Standard errors can similarly be estimated using bootstrap. This approach is identical to the two-period case, as it does not depend on the number of time periods.

\section{Practical guide}

For practitioners planning to use ML-predicted outcomes in causal analysis, I recommend the following steps.

\begin{enumerate}
    \item \textbf{Collect panel ground-truth labels.} Ensure actual outcome data is collected for a labeled subsample across at least two time periods---ideally one pre-intervention and one post-intervention period.

    \item \textbf{Train the ML model on untreated units.} Use only control units from the labeled subsample, so that the model's learned features are not contaminated by treatment exposure.

    \item \textbf{Compute the diff-vs-diff slope ($\hat{\eta}_{\epsilon}$).} For untreated units in the labeled subsample, compute $\Delta\textrm{predictedOutcome}_i$ and $\Delta\textrm{actualOutcome}_i$ across time periods. Regress the former on the latter without an intercept. The slope $\hat{\beta}$ is your estimate of $\eta_{\epsilon}$ (Section~\ref{sec:estimating_eta_epsilon}).

    \item \textbf{Use $\hat{\eta}_{\epsilon}$ to select among models.} If multiple ML architectures or feature sets are available, prefer the one with the highest $\hat{\eta}_{\epsilon}$---not the highest overall $R^2$. A high $R^2$ with low $\hat{\eta}_{\epsilon}$ is a warning sign that the model fits primarily to between-unit variation and will likely miss treatment effects.

    \item \textbf{Interpret $\hat{\eta}_{\epsilon}$ as a quality threshold.} A value near 1 means the model tracks within-unit dynamics well and is likely capable of detecting treatment effects. A value near 0 means it is not, and collecting actual outcomes for a larger share of the sample may be necessary.

    \item \textbf{Correct for attenuation (requires the stronger $\eta_T \approx \eta_{\epsilon}$ assumption).} Steps 1--5 above are useful as diagnostics and for model selection without requiring $\eta_T = \eta_{\epsilon}$. The following correction, however, does require this assumption. If $\hat{\eta}_{\epsilon} < 1$, the treatment effect estimated from ML-predicted outcomes will be attenuated by approximately that factor. Under the assumption $\eta_T \approx \eta_{\epsilon}$, an approximately unbiased estimate is:
    \begin{equation}
        \widehat{\textrm{TreatmentEffect}}_{\textrm{unbiased}} = \widehat{\textrm{TreatmentEffect}}_{\textrm{estimated}} \;/\; \hat{\eta}_{\epsilon}
    \end{equation}
    This correction should be accompanied by bootstrapped standard errors that propagate uncertainty in $\hat{\eta}_{\epsilon}$, and should be clearly flagged as resting on the $\eta_T = \eta_{\epsilon}$ assumption.
\end{enumerate}

\section{Discussion}

\subsection{Panel data requirement}
The proposed method requires panel data with at least two time periods of ground-truth outcomes for a subsample of units. Without repeated measurements, $\eta_{\epsilon}$ cannot be estimated and the diagnostic is unavailable. In settings where only cross-sectional data exist, the decomposition still provides a useful conceptual framework---it clarifies \textit{why} ML-predicted outcomes may fail to capture treatment effects---but the empirical metric cannot be computed. If experimental variation is available, $\eta_T$ can be estimated directly by comparing treatment effect estimates from predicted and actual outcomes; however, this typically requires a sample large enough to estimate the treatment effect from actual outcomes, which defeats the purpose of using ML predictions.

\subsection{Linearity}
The decomposition is presented in a linear-additive framework, but the $\eta$ coefficients can be interpreted as linear projection coefficients---the best linear approximation---regardless of the true data-generating process. The conceptual decomposition into between-unit, within-unit, and treatment-effect components does not depend on the outcome being literally linear-additive. What linearity provides is clean identification of each $\eta$ from observable moments. With nonlinear outcome processes or heterogeneous treatment effects, the $\eta$ coefficients become weighted averages, and the framework remains a useful approximation. The key conceptual insight---that a model can fit well to between-unit variation while being insensitive to treatment-induced changes---holds regardless of functional form.

\subsection{Strength of the $\eta_T \approx \eta_{\epsilon}$ assumption}
The assumption that $\eta_T \approx \eta_{\epsilon}$ is difficult to verify empirically in any given application, since verifying it requires knowing $\eta_T$, which in turn requires the experimental variation that the method is designed to avoid. The structural argument in Section 2 provides motivation: features that capture within-unit temporal variation are likely of a similar nature to features through which treatment effects operate, and both are distinct from the stable features that drive between-unit variation. However, this remains a conjecture. When in doubt, $\hat{\eta}_{\epsilon}$ should be used as a diagnostic and model-selection tool---which does not require the exact equality---rather than as a bias-correction factor, which does. Empirical validation of the relationship between $\eta_T$ and $\eta_{\epsilon}$ across different domains and data types is an important direction for future research.

\section{Conclusion}
In this paper, I have introduced a framework for decomposing ML predictions into three components: between-unit, within-unit-across-time, and counterfactual-treatment-effect. The first two components can be separately estimated with non-experimental panel data. The third one cannot be estimated absent experimental variation. Often measuring model performance using experimental variation is either impossible (because it is done before the intervention took place) or infeasible (as it would require a much bigger sample than estimating the treatment effect using collected actual outcomes directly, and so defeats the purpose of a lower cost or a larger sample). This decomposition is therefore useful when such predictions are used as outcomes in causal analysis. The key findings and contributions are:

\begin{enumerate}
\item I show that overall prediction accuracy is a poor proxy for a model's ability to detect treatment effects, as models that fit well to between-unit variation may completely miss treatment effects, especially when between-unit variation dominates.

\item I propose a metric based on within-unit variation across time ($\eta_{\epsilon}$) that better predicts a model's ability to capture treatment effects ($\eta_{T}$) under certain assumptions. I conjecture that these assumptions hold in many practical settings.

\item I demonstrate through simulations that under the assumption $\eta_{T} = \eta_{\epsilon}$, we can correct for bias in estimated treatment effects, producing unbiased estimates.

\item The approach provides a practical way to evaluate ML models for causal analysis without requiring experimental data for the entire population.
\end{enumerate}

The implications for practitioners are significant. When using ML-predicted outcomes for causal inference, researchers should not rely solely on overall prediction accuracy but should specifically evaluate the model's ability to capture within-unit variation over time. This criterion could also guide researchers in decisions about model training, for example including or excluding various features. This approach requires panel data with at least two time periods but provides valuable insight into which model is likely to perform better for causal analysis.

Future research should empirically validate the relationship between $\eta_{T}$ and $\eta_{\epsilon}$ across different domains and data types. Additionally, this work suggests that researchers might benefit from training models specifically to predict \textit{changes} rather than \textit{levels} when the ultimate goal is causal inference.

\bibliographystyle{plainnat}
\bibliography{references}

@article{blumenstock2025Togo_surveys_vs_digital_for_impact_eval,
  title = {Estimating impact with surveys versus digital traces: Evidence from randomized cash transfers in Togo},
  volume = {175},
  ISSN = {0304-3878},
  url = {http://dx.doi.org/10.1016/j.jdeveco.2025.103477},
  DOI = {10.1016/j.jdeveco.2025.103477},
  journal = {Journal of Development Economics},
  publisher = {Elsevier BV},
  author = {Aiken,  Emily and Bellue,  Suzanne and Blumenstock,  Joshua E. and Karlan,  Dean and Udry,  Christopher},
  year = {2025},
  month = jun,
  pages = {103477}
}

@article{blumenstock2025Haiti_CDR_fails_in_impact_eval,
  title = {Probing the limits of mobile phone metadata for poverty prediction and impact evaluation},
  volume = {174},
  ISSN = {0304-3878},
  url = {http://dx.doi.org/10.1016/j.jdeveco.2025.103462},
  DOI = {10.1016/j.jdeveco.2025.103462},
  journal = {Journal of Development Economics},
  publisher = {Elsevier BV},
  author = {Barriga-Cabanillas,  Oscar and Blumenstock,  Joshua E. and Lybbert,  Travis J. and Putman,  Daniel S.},
  year = {2025},
  month = may,
  pages = {103462}
}

@article{burke2021using,
  title={Using satellite imagery to understand and promote sustainable development},
  author={Burke, Marshall and Driscoll, Anne and Lobell, David B and Ermon, Stefano},
  journal={Science},
  volume={371},
  number={6535},
  pages={eabe8628},
  year={2021},
  publisher={American Association for the Advancement of Science}
}

@article{lobell2020eyes,
  title={Eyes in the sky, boots on the ground: Assessing satellite-and ground-based approaches to crop yield measurement and analysis},
  author={Lobell, David B and Azzari, George and Burke, Marshall and Gourlay, Sydney and Jin, Zhenong and Kilic, Talip and Murray, Siobhan},
  journal={American Journal of Agricultural Economics},
  volume={102},
  number={1},
  pages={202--219},
  year={2020},
  publisher={Wiley Online Library}
}

@article{Ratledge2022Uganda_satellite_electrification_and_decompression,
  title = {Using machine learning to assess the livelihood impact of electricity access},
  volume = {611},
  ISSN = {1476-4687},
  url = {http://dx.doi.org/10.1038/s41586-022-05322-8},
  DOI = {10.1038/s41586-022-05322-8},
  number = {7936},
  journal = {Nature},
  publisher = {Springer Science and Business Media LLC},
  author = {Ratledge,  Nathan and Cadamuro,  Gabe and de la Cuesta,  Brandon and Stigler,  Matthieu and Burke,  Marshall},
  year = {2022},
  month = nov,
  pages = {491–495}
}

@unpublished{Cole2025shawn_jessica_pxd_paper,
  title={The impact of digital agricultural extension service: Experimental evidence from rice farmers in India},
  author={Cole, Shawn and Goldberg, Jessica and Harigaya, Tomoko and Zhu, Jessica},
  year={2025},
  note={Working Paper}
}

@article{prentice1989surrogate,
  title={Surrogate endpoints in clinical trials: definition and operational criteria},
  author={Prentice, Ross L.},
  journal={Statistics in Medicine},
  volume={8},
  number={4},
  pages={431--440},
  year={1989},
  publisher={Wiley},
  doi={10.1002/sim.4780080407}
}

@article{frangakis2002principal,
  title={Principal stratification in causal inference},
  author={Frangakis, Constantine E. and Rubin, Donald B.},
  journal={Biometrics},
  volume={58},
  number={1},
  pages={21--29},
  year={2002},
  doi={10.1111/j.0006-341X.2002.00021.x}
}

@article{athey2025surrogate,
  title={The Surrogate Index: Combining Short-Term Proxies to Estimate Long-Term Treatment Effects More Rapidly and Precisely},
  author={Athey, Susan and Chetty, Raj and Imbens, Guido W. and Kang, Hyunseung},
  journal={The Review of Economic Studies},
  year={2025},
  publisher={Oxford University Press},
  doi={10.1093/restud/rdaf087}
}

@article{prest2023rcts,
  title={RCTs against the Machine: Can Machine Learning Prediction Methods Recover Experimental Treatment Effects?},
  author={Prest, Brian C. and Wichman, Casey J. and Palmer, Karen},
  journal={Journal of the Association of Environmental and Resource Economists},
  volume={10},
  number={5},
  pages={1231--1264},
  year={2023},
  publisher={University of Chicago Press},
  doi={10.1086/724518}
}

@misc{pettersson2025onemap,
  title={Debiasing Machine Learning Predictions for Causal Inference Without Additional Ground Truth Data: ``One Map, Many Trials'' in Satellite-Driven Poverty Analysis},
  author={Pettersson, Markus B. and Jerzak, Connor T. and Daoud, Adel},
  year={2025},
  eprint={2508.01341},
  archivePrefix={arXiv},
  primaryClass={stat.ML},
  url={https://arxiv.org/abs/2508.01341}
}

@software{reich2025prediction_decomposition,
  author = {Reich, Ofir},
  title = {Prediction Decomposition for Causal Analysis},
  year = {2025},
  url = {https://github.com/ofir-reich/prediction_decomposition_for_causal_analysis},
  version = {1.0.0}
}
\end{document}